\documentclass[conference]{IEEEtran}
\usepackage{cite}
\usepackage{amsmath,amssymb,amsfonts}
\usepackage{algorithmic}
\usepackage{graphicx}
\usepackage{textcomp}
\usepackage{xcolor}
\usepackage{multirow}

\def\BibTeX{{\rm B\kern-.05em{\sc i\kern-.025em b}\kern-.08em
    T\kern-.1667em\lower.7ex\hbox{E}\kern-.125emX}}
\begin{document}

\title{SoK: Privacy-preserving Deep Learning with Homomorphic Encryption \\

}


\author{\IEEEauthorblockN{Robert Podschwadt}
\IEEEauthorblockA{\textit{Department of Computer Science} \\
\textit{Georgia State University}\\
Atlanta, USA \\
rpodschwadt1@student.gsu.edu}
\and
\IEEEauthorblockN{Daniel Takabi}
\IEEEauthorblockA{\textit{Department of Computer Science} \\
\textit{Georgia State University}\\
Atlanta, USA \\
takabi@gsu.edu}
\and
\IEEEauthorblockN{Peizhao Hu}
\IEEEauthorblockA{\textit{Department of Computer Science} \\
\textit{Rochester Institute of Technology}\\
Rochester, USA \\
hxpvcs@rit.edu}
}

\maketitle

\begin{abstract}

Outsourced computation for neural networks allows users access to state of the art models without needing to invest in specialized hardware and know-how. The problem is that the users lose control over potentially privacy sensitive data. With homomorphic encryption (HE) computation can be performed on encrypted data without revealing its content. 
In this systematization of knowledge, we take an in-depth look at approaches that combine neural networks with HE for privacy preservation. We categorize the changes to neural network models and architectures to make them computable over HE and how these changes impact performance. We find numerous challenges to HE based privacy-preserving deep learning such as computational overhead, usability, and limitations posed by the encryption schemes.
\end{abstract}

\section{Introduction}

Neural networks(NN) have shown great success in a variety of domains and applications. Training a NN requires considerable know-how and computational resources, making a high performing model a valuable asset. This is one of the reasons for the popularity of the Machine Learning as a Service (MLaaS) model. In MLaaS a service provider (often also called server or the cloud) offers its computational resources and sometimes trained models to a client. The client, who owns the data, wants to use the computational resources and models of the service provider. This raises privacy concerns for one or both parties. Either the client needs to share their data with the server, or the server needs to share its model with the client. Solving this problem has been an active area of research, called Privacy Preserving Machine Learning (PPML). A number of approaches have been studied, relying on different techniques for privacy preservation. 

Here we will focus on solutions that are purely based on homomorphic encryption for the privacy preservation and on NN on the ML side. 
HE has been called the holy grail of encryption. It offers privacy for both the client's and the server's data. Since all of the processing can be done offline it leaks no information about the network to the client. The bulk of the NN operations can be efficiently computed using HE. This and the great popularity of NNs makes investing the combination of the two an interesting avenue of research.

HE cannot naively be applied to NNs. There are a number of challenges and limitations that need to be addressed. The exact limitations depend on the scheme used but there are common problems that arise in most schemes. 
\textbf{Multiplicative Depth (MD)} is the number of consecutive multiplications that can be applied to a ciphertext before it can no longer be decrypted correctly. It can be circumvented by using bootstrapping, but bootstrapping adds a large computational overhead. 
The \textbf{supported operations} of HE schemes are not the same as on plain data. Typically only addition and multiplication are supported, limiting the algorithms that can be run on encrypted data.
The \textbf{Complexity} of ciphertext operations is much higher than the complexity of plaintext operations both in terms of memory consumption and processing time. 

Related work has been published, however with different areas of focus. Azraoui et al. \cite{azraoui_sok_2019} also focus on neural networks and deep learning but take a broader approach to cryptographic tools, including SMC and hybrid approaches. Viand et al. \cite{viand_sok_2021} provide a systematization for FHE compilers, covering deep learning as an application for compilers, but do not include deep learning approaches that do not use a compiler. Papernot et al. \cite{papernot_sok_2018} address security and privacy issues in ML in general, whereas we focus specifically on privacy preservation. Pulido-Gaytan et al. \cite{pulido-gaytan_privacy-preserving_2021} is the most similar work to ours. They aim to provide a survey that helps researchers new to the field by giving an overview of the available libraries for ML, HE and PPML as well as a starting point for literature. We focus more on the modifications and improvements made by the individual papers, providing a more in-depth comparison and systematization of techniques used for deep learning using HE.

In this work we select 18 papers that provide a wide range of solutions to the challenges of neural networks with HE. We discuss their individual strength and weaknesses, group them systematically and in-depth based on their approaches and identify open issues and new directions for research. We further develop a set of evaluation guidelines that make it easier to compare approaches in terms of resource requirements. 
 
Our work is organized as follows: We briefly discuss the required background on HE and  neural networks in Section \ref{sec:background}. Section \ref{sec:strength_weakness} gives an overview of the approaches and their strengths and weaknesses. In Section \ref{sec:layers} we discuss how different neural network layers are handled and in Section \ref{sec:activation_functions} we take an in-depth look at activation functions for HE. We discuss general adaptions of NNs in Section \ref{sec:adaptions}. In Section \ref{sec:security} we investigate the security of the proposed systems and in Section \ref{sec:evaluation} we discuss the evaluation of the approaches.
We conclude in Section \ref{sec:challenges} by discussing challenges and future research directions.

\section{Background}\label{sec:background}

Homomorphic Encryption schemes are public key crypto schemes, which allow computation on the encrypted data. What makes HE schemes interesting is that the result of the computation is also encrypted, and no decryption is necessary during computation. Simply put, the result of computation on encrypted data is the same after decryption as if the computation was performed on plain data. HE schemes can be grouped into different categories, based on what operations they support and limitations they put on the on the circuit.

\textbf{Partially Homomorphic Encryption} (PHE) schemes are the ``simplest'' HE schemes. They only support either addition or multiplication. 

\textbf{Somewhat Homomorphic Encryption} supports both multiplication and addition on encrypted data. It cannot support arbitrarily deep circuits, making it unsuitable for some applications. 

\textbf{Leveled Homomorphic Encryption} (LHE) scheme are very similar to SHE schemes,  supporting both addition and multiplication and having a limited circuit depth. In contrast to SHE the depth needs to be controllable with a parameter $d$ during instantiation of the scheme. The size of the output of the evaluation function needs to be independent of $d$. Most modern schemes have a leveled mode.

\textbf{Fully Homomorphic Encryption} (FHE) schemes are the most powerful HE schemes. They support both addition, multiplication, and circuits of arbitrary depth. The reason circuit depth is limited in HE schemes is that, in layman's terms, the encryption adds noise to the data and the decryption process is the removal of that noise. Performing  operations on ciphertexts increases the noise and too much noise prevents correct decryption. FHE schemes circumvent this by using the bootstrapping trick \cite{gentry_fully_2009}. Boostrapping reduces the accumulated noise so that further computation can be done. This process can be repeated as many times as needed to evaluate any given circuit. Bootstrapping is computationally expensive which is why in practice many solutions do not use it. 
For more in-depth information about the different scheme types, we refer the reader to Armknect et al. \cite{armknecht_guide_nodate}. 

\subsection{HE Schemes}

Out of all the schemes used in this work the big exception is the \textbf{Paillier} scheme \cite{paillier_public-key_1999}. The scheme is only additively homomorphic. Its security is based on the decisional composite residuosity assumption. All other schemes are either leveled homomorphic or fully homomorphic.
Some of the earlier work presented here uses the \textbf{YASHE} \cite{bos_improved_2013} scheme which is based on NTRU. It has since been shown that the scheme is vulnerable to subfield lattice attacks \cite{albrecht_subfield_2016}. 
The rest of the schemes rely on the Ring Learning With Errors hardness assumption and can be grouped into three categories based on the types of plaintext they support.
\textbf{BGV} \cite{brakerski_leveled_2014} and \textbf{BFV} \cite{fan_somewhat_2012}  only support integers.
The \textbf{CKKS} or (HEAAN) \cite{cheon_homomorphic_2017} scheme supports approximate computation on real numbers. The definition of HE needs to be relaxed somewhat to include CKKS. It preforms approximate computation, therefore the result on encrypted data will differ from the result on plain data. However, the design of the scheme ensures that the approximation error happens in the least significant bits first.
\textbf{TFHE} \cite{chillotti_faster_2017} is a scheme that supports binary gates and fast bootstrapping.


\subsection{Single Instruction Multiple Data}

Some schemes support Single Instruction Multiple Data (SIMD) operations, which can offset the time needed to perform computation over encrypted data. By encoding multiple messages into a single ciphertext any operation performed on the ciphertext will be also applied to all messages encrypted in it. The number of messages that can be encoded in a ciphertext is often called slots and depends on the scheme parameters. The complexity of an operation is not dependent on how many slots are filled. SIMD operations are supported by BGV, BFV and CKKS. TFHE does not support SIMD operations.

\subsection{Neural Networks}

In this paper, we focus on privacy preservation for neural networks. Neural networks are a weighted, directed graph in which input nodes are connected to output nodes through a number of hidden nodes. Nodes, also called neurons or units, are organized in layers. 
Typically, the computation of a layer is the weighted sum of all inputs to which a non-linear activation function is applied. The weights of the network are learned during training. During training an iterative optimization algorithm, like stochastic gradient descent, reduces the loss by changing the weights.

\textbf{Neural Networks on Encrypted data}
The majority of the computation, the weighted sums, in a neural network can be performed on encrypted data. The non-linear activation functions pose the main problem. Common activation functions like ReLU, Tanh, Softmax, and Sigmoid cannot be evaluated efficiently on encrypted data. Without these non-linearities the neural network would only be able to learn linear functions.

\textbf{Training vs Inference}
Often privacy-preserving solutions for neural networks only cover inference. Training on the other hand is more computationally expensive, since it requires a forward and backward pass through the model. On top of that training requires multiple iterations of forward and backward passes. This leads to a large noise buildup, making training a much higher MD than inference. Furthermore, the weights will be encrypted with the same key as the data, meaning the server will have a model it cannot access after training is done.

\section{Strength and Weaknesses}\label{sec:strength_weakness}

In this section we give a brief overview of the selected works and their strengths and weaknesses. We group the approaches by one of their main strengths. 

\subsection{Low overhead}
Orlandi et al. \cite{orlandi_oblivious_2007} propose a privacy preserving protocol for inference using an already trained neural network. The use of a PHE scheme keeps the computational and memory overhead small. However, this forces the server to limit its part of the computation to the scalar product between the inputs and the model weights. All other steps are computed by the client.
Bourse et al. \cite{bourse_fast_2018} propose a method for privacy preserving NN inference capable of evaluating neural networks of arbitrary depth. To achieve this the authors, perform bootstrapping after every layer. Due to a greatly reduced plaintext space the used ciphertext are small however it is unclear if this approach is scalable to more complex data.

\subsection{High Throughput}
All the papers below use SIMD batching to offset some of the computational overhead, allowing them to efficiently process large batches of inputs at once. The downside is that small batches can only be processed with a large overhead.
Dowlin et al. \cite{dowlin_cryptonets_2016} propose CryptoNets one of the first solutions using FHE for neural network inference.
Chabanne et al. \cite{chabanne_privacy-preserving_2017} propose an extension of CryptoNets \cite{dowlin_cryptonets_2016}. The use of a batch normalization layer before each activation layer stabilizes training with polynomial activation functions.  
Hesamifard et al. \cite{hesamifard_cryptodl_2017} build CryptoDL a system similar to CryptoNets \cite{dowlin_cryptonets_2016}. However, the authors aim to find better low degree polynomial approximation for Sigmoid, Tanh and ReLU, thereby improving the quality of the network's predictions.
A weakness shared by all three systems is the limited plaintext space the and the accommodations that need to be made to need to be made to account for it. 
Zhang et al. \cite{zhang_encrypted_2019} develop a system for encrypted speech recognition. However, the final decoding needs to be performed by the client. 

\subsection{Improved Computation}
Lou et al. \cite{lou_she_2019} propose a system, called SHE, based on TFHE, which allows the authors to implement the ReLU function and max-pooling layers. Not needing to rely on substitutions and approximations, for those functions, removes a source of performance loss. However, TFHE is not very fast in performing matrix operations. 
Chou et al.'s \cite{chou_faster_2018} approach, called Faster CryptoNets, improves upon the work by Dowlin et al. \cite{dowlin_cryptonets_2016}, by pruning the models to reduce the number of required operations and by quantizing the weights of the network to achieve sparsity in the encoded plaintexts.
Jiang et al. \cite{jiang_secure_2018} introduce an algorithm for homomorphic encrypted matrix multiplication, based on CKKS. To multiply two $d \times d$ matrices the algorithms has a complexity of $O(d)$ using one ciphertext, which is an improvement over Halevi \cite{halevi_algorithms_2014} which has a complexity of $O(d^2)$ and needs $d$ ciphertexts. The authors build a new framework, called E2DM, for privacy preserving neural networks inference based on these algorithms. 
Brutzkus et al. \cite{brutzkus_low_2019} present multiple techniques to reduce the latency of neural networks over encrypted data, by using different data representations and ways to switch between representations. These data representations reduce the memory requirement and allow to preform operations like convolutions and matrix multiplication more efficiently. 
Mihara et al. \cite{mihara_neural_2020} present a framework for privately training neural networks. To speed up the computation the authors propose a new matrix packing technique, in which a matrix is packed diagonally into a ciphertext. This packing reduces the number of rotations and eliminates the multiplications required to transpose the matrix. A large downside of this approach is that it requires interactive phase for noise removal, despite being evaluated on a tiny dataset. 

A shared weakness of these approaches, relying on special data layout, is that the data layout needs to be designed on a case by cases basis. 

\subsection{Hardware Acceleration}
Al Badawi et al. \cite{al_badawi_towards_2020} implement and evaluate CNNs over HE ciphertexts using the BFV scheme on GPUs to speed up the computation.
For deeper networks to authors use plaintext space CRT decomposition to avoid overflows.
However, this requires frequent swapping in out of GPU memory.
PrivFT by Al Badawi et al. \cite{al_badawi_privft_2019} is a privacy preserving adaption of fasttext \cite{joulin_bag_2016} for text classification. The authors present a framework for privacy preserving inference using a plaintext model and for training an encrypted model using an encrypted dataset. To speed up inference the authors implement a Residual Number System (RNS) variant of CKKS over GPUs.

\subsection{Problems with high MD}
The following three papers find solutions for problems with large multiplicative depth.
Podschwadt \& Takabi \cite{podschwadt_classification_2020} perform privacy preserving text classification, using word embeddings and RNNs. 
However, the embedding operation is outsourced to the client, which requires sharing the embedding layer with the client. Furthermore, interaction between client and server is required to reset the noise.
Bakshi \& Last \cite{bakshi_cryptornn-privacy-preserving_2020} also investigate RNNs for privacy-preserving classification, using a network that has been trained on plaintext. However, this approach requires interactive phase for noise removal, although it only uses small datasets.   
Nandakumar et al. \cite{nandakumar_towards_2019} take on training neural networks on encrypted data. The trained model is also encrypted with the data owner's key and cannot be accessed by the server. 
All numbers in the system are encoded in their binary representation.
To keep the noise under control the authors use bootstrapping to refresh the ciphertexts. Bootstrapping is necessary after each layer in the forward and the backward pass. The downside of the approach is the large overhead brought on by bootstrapping and an inefficient data encoding.

\subsection{Useability}
Boehmer et al. \cite{boemer_ngraph-he_2019} develop ngraph-HE which is a backend for Intel's ngraph graph compiler. ngraph-HE takes existing neural networks from popular ML frameworks like TensorFlow or PyTorch and translates them into HE operations. The goal is to require as little knowledge about HE as possible. ngraph-HE can perform a number of optimizations that speed up the neural network inference. 
CHET is an optimizing compiler for tensor circuits over homomorphic encryption by Datathri et al. \cite{dathathri_chet_2019}. By using a cost model, the compiler finds the most efficient operations and crypto parameters. The optimizations that CHET performs are: encryption parameter selection, data layout selection, rotation key selection and fixed-point scaling factor selection.

\section{Neural Network Layers with HE} \label{sec:layers}

\subsection{Network architecture}

Works represented in this paper evaluate a variety of architectures. Choosing an architecture depends on the task and the data. Even on plain data it is not a simple task and requires domain knowledge and experience. Encrypted data adds another layer of complexity to choosing an architecture. Not only does the architecture need to fit the task and data, it also needs to be within the constraints of the HE scheme used. 

\textbf{Multi-layer Perceptron}
Multi-layer perceptrons or fully connected neural networks are the simplest form of neural networks. They consist of only fully connected layers. As fully connected layers are simple matrix multiplication they can be seen as the building block of all other layers. As can be seen in Table \ref{tab:supported_layers} all solutions support fully connected layers. 

\textbf{Convolutional Neural Networks (CNN)}
Common to all CNNs is the convolutional layer, but unlike fully connected networks CNNs often include other layers too. The most common are pooling layers and batch normalization layers. On plaintext the most common form pooling is max pooling. On encrypted data, the max pooling is inefficient with most schemes. The exception is SHE \cite{lou_she_2019} which uses an efficient max pooling implementation due to the TFHE scheme. Other solutions rely on various forms of average pooling.  

\textbf{Recurrent Neural Networks}
There is little work on Recurrent Neural Networks. Simple RNNs at, their core, are similar to fully connected layers. They also only consist of matrix multiplication. They tend to have higher multiplicative depth than CNNs or fully connected networks.
Podschwadt \& Takabi \cite{podschwadt_classification_2020} and
Bakashi \& Last \cite{bakshi_cryptornn-privacy-preserving_2020} use simple RNNs.
SHE \cite{lou_she_2019} is the only work to investigate LSTMs. Due to their modifications and crypto scheme, they are able to keep the multiplicative depth, that would normally come with LSTMs, under control.

\begin{table}
    \centering
    \caption{Supported Layers}
    \resizebox{\columnwidth}{!}{
    \begin{tabular}{|l|l|l|l|l|l|l|l|l|l|}
        \hline
        \textbf{Paper} & \rotatebox{90}{\textbf{Fully Connected}} & \rotatebox{90}{\textbf{Convolutions}} & \rotatebox{90}{\textbf{RNN}} & \rotatebox{90}{\textbf{LSTM}} & \rotatebox{90}{\textbf{Batch Normalization}} & \rotatebox{90}{\textbf{Max Pooling}} & \rotatebox{90}{\textbf{Average Pooling}} & \rotatebox{90}{\textbf{Scaled Average Pooling}} \\
        \hline
        Orlandi et al. \cite{orlandi_oblivious_2007} & 
         $\bullet$ & - & - & - & - & - & - & -  \\
        \hline
        Al Badawi et al. \cite{al_badawi_towards_2020}     & 
         $\bullet$ & $\bullet$ & - & - & - & - & $\bullet$ & -  \\ 
        \hline
        Nandakumar et al. \cite{nandakumar_towards_2019}   & 
         $\bullet$ & - & - & - & - & - & - & - \\
        \hline
        Bourse et al. \cite{bourse_fast_2018}       & 
         $\bullet$ & - & - & - & - & - & - & - \\
        \hline
        Chabanne et al. \cite{chabanne_privacy-preserving_2017} & 
         $\bullet$ & $\bullet$ & - & - & $\bullet$ & - & $\bullet$ & - \\
        \hline
        Faster Cryptonets \cite{chou_faster_2018} & 
         $\bullet$ & $\bullet$ & - & - & $\bullet$ & - & - & $\bullet$ \\
        \hline
        CHET \cite{dathathri_chet_2019} & 
         $\bullet$ & $\bullet$ & - & - & - & - & $\bullet$ & - \\
        \hline
        Cryptonets \cite{dowlin_cryptonets_2016} & 
         $\bullet$ & $\bullet$ & - & - & - & - & $\bullet$ & $\bullet$ \\
        \hline
        CryptoDL \cite{hesamifard_cryptodl_2017} & 
         $\bullet$ & $\bullet$ & - & - & $\bullet$ & - & - & $\bullet$ \\
        \hline
        E2DM \cite{jiang_secure_2018}  &
         $\bullet$ & $\bullet$ & - & - & - & - & $\bullet$ & - \\
        \hline
        SHE \cite{lou_she_2019} &
         $\bullet$ & $\bullet$ & - & $\bullet$ & $\bullet$ & $\bullet$ & - & - \\
        \hline
        Brutzkus et al. \cite{brutzkus_low_2019} & 
         $\bullet$ & $\bullet$ & - & - & - & - & $\bullet$ & - \\
        \hline 
        ngraph-he \cite{boemer_ngraph-he_2019} & 
         $\bullet$ & $\bullet$ & - & - & $\bullet$ & - & - & $\bullet$ \\
        \hline 
        Mihara et al. \cite{mihara_neural_2020} & 
         $\bullet$ & - & - & - & - & - & - & - \\
        \hline
        Podschwadt \& Takabi \cite{podschwadt_classification_2020} & 
         $\bullet$ & - & $\bullet$ & - & - & - & - & - \\
        \hline
        PrivFT \cite{al_badawi_privft_2019} & 
         $\bullet$ & - & - & - & - & - & - & - \\
        \hline
        CryptoRNN \cite{bakshi_cryptornn-privacy-preserving_2020} & 
         $\bullet$ & - & $\bullet$ & - & - & - & - & - \\
        \hline
        Zhang et al. \cite{zhang_encrypted_2019} & 
         $\bullet$ & - & - & - & - & - & - & - \\
        \hline
    \end{tabular}
    }
    \label{tab:supported_layers}
\end{table}


\subsection{Fully Connected and Convolutional Layers } \label{sec:fc_conv_layers}
Fundamentally, fully connected and convolutional layers are similar. Both rely on matrix multiplications and dot products. Matrix multiplication and dot products are HE compatible since they only consist of additions and multiplications. No work needs to be done to adapt them to HE. However, the naive implementation can be quite costly.
In the naive implementation every matrix or vector element is a single ciphertext. The naive implementation of $d$-dimensional matrix  multiplication requires $d^3$ multiplications and the dot product of two $d$-dimensional vectors requires $d$ multiplications. 

One strategy for speeding up computation is to reduce the number of operations needed.
SHE \cite{lou_she_2019} cuts down on the number of multiplications that are required during the matrix multiplication and dot product by replacing them with shift operations. This is possible due the quantization of the network weights to be a power two and the network inputs being represented as fix point numbers in their binary representation. Faster CryptoNets \cite{chou_faster_2018} uses a very similar approach. The authors also quantize the network weights to be a power of two giving them a sparse polynomial representation, mononomial in fact. Multiplication between a ciphertext and these sparse plaintexts is much faster. Research on plaintext \cite{zhou_incremental_2017} suggests that quantization of the weights can be done without sacrificing accuracy.

Another option is using the SIMD operations offered by many schemes. The goals are to use fewer ciphertexts, decrease latency and to organize the data in such a way that it speeds up computation. This is called ciphertext packing. Cleverly packed ciphertexts can drastically reduce the number of operations needed to evaluate specific algorithms. 
E2DM \cite{jiang_secure_2018} makes use of an efficient matrix multiplication algorithm. The authors define an encoding map that allows the encoding of one or more matrices in single ciphertext and subsequently enables them to perform efficient matrix multiplications. 
PrivFT \cite{al_badawi_privft_2019} uses a packing structure that packs the input in a relatively small number of ciphertexts. For efficient dot products the embedding matrix is packed vertically.
Brutzkus et al. \cite{brutzkus_low_2019} introduce different ways of encoding the data into ciphertexts, allowing them to perform certain operations, such as convolutions and matrix multiplications, more efficiently. Throughout processing, the representations can be switched as needed. These encodings use fewer ciphertexts and are therefore much more memory efficient. 
CHET \cite{dathathri_chet_2019} also supports a number of different data layouts but it also has cost models associated with each. Based on the input model, selected scheme, and input data, CHET can select the best data layout and packing strategy. Whereas in other approaches the layout selection needs to be done by hand.
Mihara et al. \cite{mihara_neural_2020} use a diagonal ciphertext packing that allows for efficiently transposing a matrix which is important during the backpropagation in training. 

The main difficulty in using ciphertext packing is that the packing strategy needs to be designed for the dimensionality of the data and the algorithm. There is no one size fits all solution. Without compilers like CHET \cite{dathathri_chet_2019} finding a good packing strategy can require quite a bit of trial and error. Furthermore, packing often requires rotations of the slots. The rotations need to be taken into account when calculating the noise budget. The same goes for the transformation from representation to another. These also need to be taken into account in the noise budget and run time. 
SIMD operations are supported by most schemes. TFHE does not support them and therefore cannot make use of these packing strategies.

\subsection{Pooling Layers}

Pooling layers are often crucial to the success of CNNs. On plaintext, max pooling is very popular. But in the encrypted domain it is infeasible most of the time due to impracticality of comparison. The exception is SHE \cite{lou_she_2019}. It can actually evaluate max pooling due to the binary representation of the data and the use of the TFHE. 
Some work replaces the maxpooling operation with either average pooling or scaled average pooling. Some papers \cite{chabanne_privacy-preserving_2017, dathathri_chet_2019, al_badawi_towards_2020} actually use average pooling, while others use scaled average pooling \cite{hesamifard_cryptodl_2017, zhang_encrypted_2019, dowlin_cryptonets_2016, chou_faster_2018, boemer_ngraph-he_2019}.
Scaled average pooling is average pooling scaled by a constant factor. The most common scaling factor is the number of input elements, this way scaled average pooling turns into a sum of the inputs. The advantage of using average pooling is the tighter control over the magnitude of the values. The parameters of the crypto scheme need to be chosen so that all values during computation fit into the plaintext space. With average pooling the output of the pooling operations are in the same range as the inputs. With scaled average pooling the output values can grow larger. The cost of using average pooling is an additional multiplication. 

\subsection{Batch Normalization Layers}

On plain data, batch normalization is used to stabilize the training process by reducing the internal covariate shift. This is done by forcing the inputs to a layer to follow a zero mean normal distribution. This normalizing operation makes inputs with large absolute value less likely. This effect was first used by Chabanne et al. \cite{chabanne_privacy-preserving_2017} to stabilize training with polynomial activation functions. Polynomial activations usually work well in a small interval around zero. Outside of that interval they are unbounded and have rapidly growing derivatives, which cause problems during training. By placing a batch normalization before a polynomial activation layer, the probability of encountering values outside of the optimal range of polynomial activation decreases. Batch normalization has been incorporated into other works \cite{hesamifard_cryptodl_2017, zhang_encrypted_2019, chou_faster_2018, lou_she_2019, boemer_ngraph-he_2019} as well. In addition to stabilizing training, it also helps preventing values from overflowing the limits of the plaintext space.

\subsection{Recurrent Layers}

Recurrent layers pose a problem for encrypted execution. The MD of an RNN is usually much larger than that of a feed-forward network with a similar number of parameters. The depth of an RNN depends on the number of elements in the input sequence. For a sequence with $n$ elements, it is equivalent to a fully connected network with $n$ layers. If the same activation function is used. Typically, RNNs use the Tanh function which requires a polynomial of degree 3 to be adequately approximated. LSTMs, which normally offer better performance, have an even larger multiplicative depth due to their more complex internal structure. Dealing with large multiplicative depth requirements is the main challenge of adapting RNNs for encrypted data.   
SHE \cite{lou_she_2019} tackles the multiplicative depth problem in two ways. Firstly, the authors replace all the multiplications with shift operations, which can be performed at no cost to the noise budget. Additionally, the authors replace the activation functions in the LSTM cells with ReLU, which is easier to compute than the standard Tanh function.
Podschwadt \& Takabi \cite{podschwadt_classification_2020} implement simple RNNs using naive matrix multiplication and a degree three polynomial Tanh approximation. The noise buildup prevents the network from completely processing the inputs. The authors measure the remaining noise budget and use the client for interactive noise removal. Using interactive phases to refresh the noise is also used by CryptoRNN \cite{bakshi_cryptornn-privacy-preserving_2020}. CryptoRNN also uses simple RNNs, but they refresh noise at different points in the network than Podschwadt \& Takabi \cite{podschwadt_classification_2020}. They propose three different points at which the data is sent to the client for noise refresh; 1) after every multiplication, 2) sending the data to the client to evaluate the non-linear activation function, 3) refreshing the noise after processing each sequence element. 
Standard RNN architectures are not well suited for execution over homomorphically encrypted data. Either fundamental changes need to be made as in \cite{lou_she_2019} or an interactive protocol needs to be used. However, the interactive protocol takes away some of the key advantages of HE over SMC which is the lower communication overhead and the ability to perform the computation independently from the client. The great strength of RNNs is that they have some ``memory'' of previous states. But on encrypted data it is their downfall since this ``memory'' produces networks with a large multiplicative depth.

\subsection{Embedding Layers}
Word embeddings are a common tool in natural language processing. They are a real valued vector representation of a word's meaning. Word embeddings are interesting because they make good features for neural networks. The process of arriving at the vector representation differs from method to method. 
Word embeddings are used by two works in this paper. The first one, PrivFT \cite{al_badawi_privft_2019} relies on the fasttext architecture. For processing by the network, the words need to be encoded into a one-hot vector representations. The one-hot encoding, in modified form, is done by the client. Normally the translation from a word to one hot encoding would happen on the server side. However, this operation is prohibitively costly over encrypted data. The client sums up all the one-hot encodings, encrypts the results and sends the ciphertext to the server, alongside the number of words that went into the encoding. The number of words is sent in plain. On the server the vector is multiplied with the embedding matrix and the result is scaled by the number of words that were used to create the vector by the client. Podschwadt \& Takabi \cite{podschwadt_classification_2020} also work with textual data. In their work they outsource the embedding to the client. The client performs the embedding operation, which is a table lookup, and sends the encrypted result to the server for computation. 
In both approaches some information about the model and the data needs to be shared with the client. Both need to provide the client with an enumerated vocabulary of all the words known to the model. This only leaks very little information. Podschwadt \& Takabi \cite{podschwadt_classification_2020} also requires the client to have access to the learned embeddings in plaintext. This is not a problem if publicly available pretrained embeddings are used. But it prevents the server from using fine-tuned or self-learned embeddings and keeping them secret from the client. Another issue is that embeddings further increase the size of the data. A single word is often turned into vector of 100+ real values, which further increases in size during encryption.

\section{Activation Functions}\label{sec:activation_functions}

The types of activation functions used for privacy-preserving ML range from simple to complex. Privacy-preserving approaches to ML are necessarily different from standard ML in their choice of activation function. Low degree polynomials are a popular choice for activation functions since most schemes can easily evaluate polynomials.

\textbf{ReLU:}
The simplest non-linear polynomial, the square function $f(x)=x^2$, is often used as a replacement for ReLU \cite{dowlin_cryptonets_2016, zhang_encrypted_2019, bakshi_cryptornn-privacy-preserving_2020, al_badawi_towards_2020, jiang_secure_2018, mihara_neural_2020, boemer_ngraph-he_2019, brutzkus_low_2019, hesamifard_cryptodl_2017}.
The square function has the benefit that it is fast to compute and adds little overhead to the computation. However, it does not perform well in all problem domains. For example, it cannot be used in RNNs due to its rapidly growing derivative \cite{podschwadt_classification_2020}. 
Chabanne et al. \cite{chabanne_privacy-preserving_2017} use a more complex polynomial approximation of the Relu function. 
Lou et al. \cite{lou_she_2019} use TFHE scheme which enables the authors to use  ReLU on encrypted data. 

\textbf{Sigmoid:}
Orlandi et al. \cite{orlandi_oblivious_2007} use the sigmoid activation function, but the evaluation is done by the client. The server sends the encrypted inputs to the client. The client decrypts the values, applies the activation, encrypts the results and sends it back to the server.
Zhang et al. \cite{zhang_encrypted_2019} replace the sigmoid function with $-1/48x^3+1/4x+1/2$.
Nandakumar et al. \cite{nandakumar_towards_2019} do not use polynomial approximations. Instead, they use a lookup table to compute the sigmoid function. The client needs to precompute the table prior to running the network.

\textbf{Tanh:}
Tanh is often used in RNNs. Both works CryptoRNN \cite{bakshi_cryptornn-privacy-preserving_2020} and Podschwadt \& Takabi \cite{podschwadt_classification_2020} use it. Hesamifard et al. \cite{hesamifard_cryptodl_2017} find an approximation for Tanh but do not use it in a network. However, their approach is used in \cite{podschwadt_classification_2020}. Both Tanh and Sigmoid approximations need to be of at least degree three to capture the negative and positive output values, but this comes at the cost of a larger MD. When designing solutions lower degree polynomials should be tried first. 

\textbf{Softmax:}
Al Badawi et al. \cite{al_badawi_privft_2019}, unlike most other solutions, use $1/8x^2+1/2x+1/4$ as an approximation of the softmax function. Since the softmax function is usually the last operation in a neural network most other works leave its evaluation to the client. This poses a risk however, since directly exposing the logits of the neural network to the client makes it more vulnerable to adversarial and model inversion attacks.

\subsection{Finding Polynomial Activation Functions}

Using polynomial activation is an attractive alternative to the functions used on plaintexts. Most schemes can evaluate them easily and multiple approaches have been proposed to find good approximations. Using polynomials in training is difficult since their unbounded derivatives can lead to exploding gradients. \cite{chabanne_privacy-preserving_2017} addresses this issue by combining polynomial activations with batch normalization.

\cite{chabanne_privacy-preserving_2017} use polynomial regression to find approximate activation functions.
Hesamifard et al.  \cite{hesamifard_cryptodl_2017} approximate activation functions based on their derivative. 
CHET \cite{dathathri_chet_2019} and ngraph-he \cite{boemer_ngraph-he_2019} use a different approach. 
The authors also use degree two polynomials of the form $f(x)=ax^2+bx$. But rather than finding an approximation the values $a$ and $b$ are learned during training of the network.

\subsection{Interactive Activation Functions}
CryptoRNN \cite{bakshi_cryptornn-privacy-preserving_2020} also investigates interactive phases at various points in the computation. One option presented is to outsource the computation of the nonlinear activation to the client. 
Orlandi et al. \cite{orlandi_oblivious_2007}, due to the limitations Paillier scheme, the client applies the activation function after performing the decryption.

\subsection{Others}
Bourse et al. \cite{bourse_fast_2018} use a step function which outputs either $1$ or $-1$ based on the sign of the input value.
Nandakumar et al. \cite{nandakumar_towards_2019}, instead of using an approximate activation, use the lookup table approach described by Crawford et al. \cite{crawford_doing_2018}, for the Sigmoid function.

\section{Adaptions to HE constraints} \label{sec:adaptions}

\subsection{Data encoding}

Integers are the most commonly supported plaintexts. Real numbers or only supported by CKKS and TFHE only supports individual bits. Most plaintext ML applications work on either 32 or 64 bit floating point numbers. This means that solutions using CKKS, \cite{podschwadt_classification_2020, bakshi_cryptornn-privacy-preserving_2020,mihara_neural_2020,al_badawi_privft_2019, jiang_secure_2018} can mostly just use real numbers. But since computation in CKKS is approximate these solutions need to make sure to choose the precision parameter correctly. Not enough precision can change the classification on encrypted data. 

Solutions based on other schemes need to adapt to the reduced plaintext space. CryptoDL \cite{hesamifard_cryptodl_2017} choose an only integer solution. They scale the weights of the network and round them to be integers. Al Badawi et al. \cite{al_badawi_towards_2020} use a similar approach. Their solution also uses integers as inputs but thy use CRT to decompose the values into multiple parts. This decomposition keeps individual values smaller and thereby leaves room for intermediate values. Large intermediate values can become a problem in CryptoDL \cite{hesamifard_cryptodl_2017}. The authors need to make sure that intermediate values never overflow the plaintext space by choosing the crypto parameters large enough.

Fixed point encoding is used by many solutions \cite{dathathri_chet_2019, dowlin_cryptonets_2016,zhang_encrypted_2019,chou_faster_2018, chabanne_privacy-preserving_2017, lou_she_2019, nandakumar_towards_2019} to address the limit plaintext space. Fixed point encoding suffers from the same problems that other integer only solutions suffer from, mainly the overflow of the plaintext space. 

SHE \cite{lou_she_2019} and Nandakumar et al. \cite{nandakumar_towards_2019}, despite using different encryption schemes, encode the fixed-point number in their binary representation. In the case of SHE \cite{lou_she_2019} this is due to the limitation of the encryption scheme. Nandakumar et al. \cite{nandakumar_towards_2019} however use the binary representation to use a lookup table as activation functions. This binary representation however leaves a lot of the ciphertext's capacity unused, as it can hold much larger numbers than a single bit. 

Bourse et al. \cite{bourse_fast_2018} have an even more radical encoding approach. The authors map all input values to either -1 or 1 to fit their computation scheme. They do this by dividing the original message space in half and map everything equal or large than the mean to 1 and everything smaller to -1. This loses a lot of information. While all other schemes have to deal with small losses due to their encoding or approximate computation, here the authors reduce the information by over 99\%. In our opinion this is only possible due MNIST being used in the experiments. The MNIST images are already high contrast, grey scale and have very little color gradients. Removing 7 of the 8 bits of information should not harm the recognizability of the digits. The authors note: ``...a quick visual inspection of the result shows that the digits depicted in the images are still clearly recognizable.'' It is questionable if this would hold for other datasets than MNIST.

\subsection{Weight Constraints and Conversion}

Some solutions place further constraints on the weights or their encoding. This is often done to speed up computation. As discussed in the previous section solutions that work on integers or fixed-point encoding  \cite{dathathri_chet_2019, dowlin_cryptonets_2016, chabanne_privacy-preserving_2017, lou_she_2019, nandakumar_towards_2019, al_badawi_towards_2020, hesamifard_cryptodl_2017} need to transform their weights into integer as well. This is done by applying a scaling factor to the weight of each layer. The scaling factor needs to take the non-linear activation functions into consideration. This can cause the weights to grow quite large in the later layers of the network causing overflow problems.

Other solutions use varying forms of quantization.
In \cite{chou_faster_2018} the authors use quantization to enforce sparsity in the polynomial representation of the weights, speeding up the computation. Zhang et al. \cite{zhang_encrypted_2019} use Lloyed-max quantization the networks weights during training. This is done to increase network performance on fixed point numbers and not to speed up computation like in other solutions.
SHE \cite{lou_she_2019} uses fixed point numbers in their binary representation. To speed up neural network computation, the authors use log-quantization to transform all weights to be powers of 2. This allows them to replace the multiplications of the weights with shifts and accumulators. Shifting operations are very fast since they do not require any operation on the encrypted data itself.
Bourse et al. \cite{bourse_fast_2018} discretize the weights to be either -1 or 1. 

\subsection{Constant Folding}

As discussed earlier reducing the number of operations increases performance. A number of solutions include constant folding as way to reduce the number of operations. The goal is to combine the constant factors of successive operations to into one, so it has to be applied only once.
Boehmer et al. \cite{boemer_ngraph-he_2019} describe how to fold some factors of the activation into previous layers weights. When a convolutional or fully connected layer is followed by a polynomial activation layer one of the polynomial coefficients can be moved to the previous layer's weights, removing one multiplication. They further describe how to fold a batch normalization layer into the previous layer.
Dowlin et al. \cite{dowlin_cryptonets_2016} mention that an average pooling layer and a linear layer, e.g. convolutional layer, can be collapsed or folded. This collapse is described in more detail in \cite{boemer_ngraph-he_2019}. An average pooling layer followed by a convolutional layer can be replaced by a scaled mean pooling layer and a convolutional layer by moving the scaling factor into the weights of the linear layer.

\subsection{Client-side Computation}

To work around some of the limitations of HE, some solutions offload work to the client. 
Chou et al. \cite{chou_faster_2018} and Brutzkus et al. \cite{brutzkus_low_2019} take a transfer learning like approach. Chou et al. \cite{chou_faster_2018} use the topmost layers of a network already trained on a general task and fine tune the lower layers on a specific task. The upper layers are sent to and evaluated by the client on plaintext. The output of these layers is encrypted and sent to the server where the lower layers are evaluated. This approach places a larger part of the computation on the client. Brutzkus et al. \cite{brutzkus_low_2019} follow a very similar approach where the client runs data through a pre-trained network. The output of the network is encrypted and sent to the server for further processing. This output is smaller than the original inputs and can be classified using fairly shallow networks. The client-side network works as a form of dimensionality reduction. Both solutions place a lot of the computational burden on the client.
Podschwadt \& Takabi \cite{podschwadt_classification_2020} use a somewhat similar approach in which they send the embedding part of the network to the client and the client performs the embedding operation. Compared to the other two solutions this is less computational expensive for the client but requires a larger data transfer. 

CryptoRNN \cite{bakshi_cryptornn-privacy-preserving_2020} and Orlandi et al. \cite{orlandi_oblivious_2007} go even further and have the client compute the activation function. 
These approaches, in which the client performs large parts of the computation, do not fit as well into the MLaaS model. At least not if the goal is to alleviate the client of computational work. The goal here is for both sides to protect the privacy of their data, the model on the server side and the network inputs on the client side.

\subsection{Interactive Phases}

Despite one of the main advantages of HE is that computation can be performed offline, a number of solutions include interactive phases for various purposes. Offline computation in this setting means that there is no communication required beyond the initial transfer of the data. 
In Orlandi et al. \cite{orlandi_oblivious_2007} the server only computes scalar products, the client performs the rest of computation, including the activation function. CryptoRNN \cite{bakshi_cryptornn-privacy-preserving_2020} use a similar approach but not to the same extend. The server sends the preactivation values to the client where they are evaluated on plaintext and sent back to server in encrypted form. In the other scenarios the authors consider they use the client to refresh the noise in the ciphertexts. The noise refresh can happen at different points in the computation. In their works the authors evaluate three possible points. After every multiplication, after every activation function, or after computation of a single sequence element. Podschwadt \& Takabi \cite{podschwadt_classification_2020} also uses the client for noise removal. The authors monitor the noise in the ciphertext perform and interactive noise removal when necessary. Mihara et al. \cite{mihara_neural_2020}, in their training setup, use the client to remove the noise from the data after every minibatch.

\subsection{Batching}

In Section \ref{sec:fc_conv_layers} we already discussed ciphertext packing as a use of SIMD processing. Another way of using SIMD processing is to process multiple instances at once, at no additional cost. This is called batching. A common form of batching is to encrypt the same feature of multiple instances into a single ciphertext, leading to one ciphertext per feature. Since ciphertexts are usually orders of magnitude larger than plaintexts this form of batching requires a lot of memory. It is also less efficient in terms of single instance or small batch processing. Supported batch sizes are usually large ($>1000$) and the memory and runtime requirements are independent of the number of instances in a batch. It makes no difference if 1 or 1000 instances are being processed. These systems typically have decent amortized per instance time, but the latency is rather larger, making them more attractive for bulk processing and less attractive for close to real-time applications. CryptoNets \cite{dowlin_cryptonets_2016} uses this approach with a batch size of 4096. CryptoDL \cite{hesamifard_cryptodl_2017} and Al Badawi et al. \cite{al_badawi_towards_2020} support a batch size of 8192. Podschwadt \& Takabi \cite{podschwadt_classification_2020} use a comparatively small batch size of 128 for their works on RNNs. Nandakumar et al. \cite{nandakumar_towards_2019} also work with smaller batch sizes of 60 or 1800 instances depending on the parameters. 

\subsection{Training on encrypted data}


Even with powerful computational resources available training needs to be severly constraint to be somewhat practical. It either needs interaction with the client as used by Mihara et al. \cite{mihara_neural_2020}
or the task needs to be simplified. Typically, on plain data, model train for many epochs. \cite{al_badawi_privft_2019} needs to constrain the training to two epochs with large minibatch sizes. Nandakumar et al. \cite{nandakumar_towards_2019} circumvent this problem by using bootstrapping. However, bootstrapping is so expensive that the authors need to drastically simplify the problem. The model the authors train is quite small and the input data is subsampled from 28x28 to 8x8. Even with these simplification training one minibatch takes 40 minutes. With the parameter settings used in the paper a minibatch contains 60 instances. The MNIST dataset contains 60,000 training instances. This means one epoch of training would take 666.7 hours or almost 28 days. 

\section{Security} \label{sec:security}

\subsection{Encryption Schemes}

The choice of encryption scheme has a number of consequences. 1) It determines what operations are available and thereby what type of activation and architectures can be used. 2) It determines the plaintext space. Messages need to be encoded into the plaintext space. Most schemes, BGV, BFV, YASHE and FV, only support integers. CKKS supports real numbers and TFHE only supports individual bits. 3) The encoding procedure is also influenced by the scheme's support for SIMD operations. 

\cite{dowlin_cryptonets_2016, zhang_encrypted_2019} use the YASHE \cite{bos_improved_2013} crypto scheme, which should not be used anymore. Orlandi et al. \cite{orlandi_oblivious_2007} use the Paillier crypto scheme. Unlike the other systems this system is only additively homomorphic.
Very common schemes are BGV and BFV. Both schemes share a lot of similarities such as SIMD operations and integer only message space. BGV is used by Chabanne et al. \cite{chabanne_privacy-preserving_2017}, CryptoDL \cite{hesamifard_cryptodl_2017}, 
and Nandakumar \cite{nandakumar_towards_2019}. Al Badawi et al. \cite{al_badawi_towards_2020} use the BFV scheme. Faster Cryptonets \cite{chou_faster_2018} uses an RNS variant of BFV scheme \cite{bajard_full_2016}. \cite{nandakumar_towards_2019} is the only work relying on BFV or BGV that uses bootstrapping. 

Another popular scheme is CKKS. The obvious advantage of CKKS is the native support of real numbers as the message space. In BFV/BGV real numbers need to be simulated by using a fixed-point encoding. CHET \cite{dathathri_chet_2019}, E2DM \cite{jiang_secure_2018}, CryptoRNN  \cite{bakshi_cryptornn-privacy-preserving_2020}, and Podschwadt \& Takabi \cite{podschwadt_classification_2020} use CKKS. PrivtFT \cite{al_badawi_privft_2019} also uses CKKS but in an RNS variant. While CKKS supports bootstrapping none of the above solutions use it. 

In contrast to the prior schemes, bootstrapping in TFHE is faster. However, it only supports single bits as the message space. This paired with the fact that it does not support SIMD operations makes it less attractive for performing matrix multiplications. Bourse et al. \cite{bourse_fast_2018} makes up for this fact by binarizing the entire network and thereby reducing the complexity of operations.
SHE \cite{lou_she_2019} quantize the network weights and replace the expensive multiplications with much faster shift operations.  

\subsection{Crypto Library}

The most widely used library is Microsoft's SEAL \cite{noauthor_microsoft_2020}. Especially more recent publications seem to favor SEAL. It offers support for the BFV and CKKS scheme. Seal is used by \cite{dowlin_cryptonets_2016, chou_faster_2018, bakshi_cryptornn-privacy-preserving_2020,dathathri_chet_2019,al_badawi_privft_2019,al_badawi_towards_2020}.
HELib \cite{halevi_algorithms_2014}, one of the early FHE libraries to come out, supports the BGV and CKKS scheme. It is used by \cite{chabanne_privacy-preserving_2017, hesamifard_cryptodl_2017, podschwadt_classification_2020, nandakumar_towards_2019}
The HEAAN library is the only library represented in this work that supports CKKS bootstrapping. It only supports the CKKS scheme and is developed by the original authors of the CKKS paper. HEAAN is used by \cite{jiang_secure_2018,dathathri_chet_2019}. 
The authors behind the TFHE scheme also offer a library implementing the scheme by the same name. The TFHE library \cite{chillotti_tfhe_2020} is used by \cite{bourse_fast_2018, lou_she_2019}. 
Al Badawi et al. \cite{al_badawi_privft_2019, al_badawi_towards_2020} also use another library named A*HE \cite{al_badawi_high-performance_2018}. This library allows to accelerate HE using GPUs. The individual operations like homomorphic addition, multiplication, etc. are sped up by factors between 10 and 250 depending on the operation and the crypto parameters. 

\begin{table}
    \centering
    \caption{Resource Requirements. *GPU memory; System main memory unknown. **Total system memory; upper bound}
    \resizebox{\columnwidth}{!}{
    \begin{tabular}{|c|c|c|c|r|r|}
        \hline
        Paper                                                       & Train.  & Inf. & Dataset   & Memory & Communication\\
        \hline
        \multirow{2}{*}{Al Badawi \cite{al_badawi_towards_2020}}    &           & $\bullet$ & MNIST     & $<16$* GB   &  0.98 GB           \\
                                                                    &           & $\bullet$ & CIFAR-10  & $<16$* GB   &  1.77 GB            \\
        \hline
        Nandakumar \cite{nandakumar_towards_2019}                   & $\bullet$ & $\bullet$ & MNIST     & $<250$** GB &  -                 \\
        \hline
        Bourse et al. \cite{bourse_fast_2018}                       &           & $\bullet$ & MNIST     &  -            & $7.82$ KB \\
        \hline
        Chabanne et al. \cite{chabanne_privacy-preserving_2017}     &           & $\bullet$ & MNIST     & -             &  -                \\
        \hline
        \multirow{3}{*}{Faster Cryptonets \cite{chou_faster_2018}}  &           & $\bullet$ & MNIST     & $<48$**  GB & 0.41           \\
                                                                    &           & $\bullet$ & CIFAR-10  & $<1433$** GB &  1.57              \\
                                                                    &           & $\bullet$ & \cite{gulshan_development_2016} & $<1433$**   &  1183.8 GB              \\
        \hline
        \multirow{2}{*}{CHET \cite{dathathri_chet_2019}}            &           & $\bullet$ & MNIST     & $<224$** GB & -                 \\
                                                                    &           & $\bullet$ & CIFAR-10  & $<224$** GB & -                 \\
        \hline
        Cryptonets \cite{dowlin_cryptonets_2016}                    &           & $\bullet$ & MNIST     & $<16$** GB  & 0.36 GB              \\
        \hline
        \multirow{2}{*}{CryptoDL \cite{hesamifard_cryptodl_2017}}   &           & $\bullet$ & MNIST     & $<16$** GB  & -                 \\
                                                                    &           & $\bullet$ & CIFAR-10  & $<16$** GB    & -                 \\
        \hline
        E2DM \cite{jiang_secure_2018}                               &           & $\bullet$ & MNIST     &  -            &  0.02 GB             \\
        \hline
        \multirow{4}{*}{SHE \cite{lou_she_2019}}                    &           & $\bullet$ & MNIST     & $<1024$** GB  & 0.12 GB             \\
                                                                    &           & $\bullet$ & CIFAR-10  & $<1024$** GB  & 0.16 GB               \\
                                                                    &           & $\bullet$ & ImageNet  & $<1024$** GB  & 7.7 GB                \\
                                                                    &           & $\bullet$ & Penn Treebank & $<1024$** GB & 0.01 GB                \\
        \hline
        \multirow{2}{*}{Brutzkus et al. \cite{brutzkus_low_2019}}              &           & $\bullet$ & MNSIT     &  -            &  -                \\
                                                                    &           & $\bullet$ & CIFAR-10  &  12           &  -                \\
        \hline
        \multirow{2}{*}{ngraph-he \cite{boemer_ngraph-he_2019}}     &           & $\bullet$  & MNSIT     & $<376$** GB   & -                 \\
                                                                    &           & $\bullet$  & CIFAR-10  & $<376$** GB  & -                 \\
        \hline
        Mihara et al. \cite{mihara_neural_2020}                     & $\bullet$ & $\bullet$ & Iris      & -             & -                 \\
        \hline
        Podschwadt \& Takabi \cite{podschwadt_classification_2020}  &           & $\bullet$ & IMDB      & $<32$** GB              &  15.20 GB                \\
        \hline
        \multirow{5}{*}{PrivFT \cite{al_badawi_privft_2019}}        &           & $\bullet$ & IMDB      & $<16$* GB      &   0.38 GB               \\
                                                                    &           & $\bullet$ & YELP      & $<16$* GB     &   0.38 GB                \\
                                                                    &           & $\bullet$ & AGNews    & $<16$* GB     &   0.38 GB               \\
                                                                    &           & $\bullet$ & DBPedia   & $<16$* GB     &   0.38 GB               \\
                                                                    & $\bullet$ & $\bullet$ & YouTube Spam   &  120 GB     &   549 GB               \\
        \hline
        \multirow{4}{*}{CryptoRNN \cite{bakshi_cryptornn-privacy-preserving_2020}}  &              & $\bullet$ & \cite{ananda_freire_wall-following_nodate}    & -                                                                                 & $1.6$ MB \\
                                                                                    &              & $\bullet$ & \cite{torres_sensor_2013}                     & -         &   $2.7$ MB    \\
                                                                                    &              & $\bullet$ & \cite{oliver_roesler_eeg_nodate}              & -           &  $1.2$ MB  \\
                                                                                    &              & $\bullet$ & \cite{candanedo_accurate_2016}                & -           &  $9.5$ MB \\
        \hline
        \multirow{2}{*}{Encrypted Speech \cite{zhang_encrypted_2019}}&              & $\bullet$ & Switchboard  & -             &  -                \\
                                                                     &              & $\bullet$ & Cortana      & -             &  -                \\
        \hline
        Orlandi et al \cite{orlandi_oblivious_2007}                 &               & $\bullet$ & \cite{gorman_analysis_1988}   &   -            & -                  \\
        \hline

    \end{tabular}
    }
    \label{tab:ressources}
\end{table}


\subsection{Information Leakage}

In Orlandi et al. \cite{orlandi_oblivious_2007}  the server obfuscates the results it returns to the client in order to prevent the client from learning the model parameters. Furthermore, to protect the model topology the authors add fake nodes into the network. Their only purpose is to hide the architecture from the client. 

CryptoRNn\cite{bakshi_cryptornn-privacy-preserving_2020} and Podschwadt \& Takabi \cite{podschwadt_classification_2020} face the problem of revealing internal states to the client during noise removal. This could allow the client to learn the parameters of the model. This can be circumvented by adding some randomness to the data before sending it to the client for noise removal. The randomness can be removed once the reduced noise ciphertext is returned to the server. This obfuscation becomes more complicated when the client is used for computing non-linear functions as in CrypotRNN \cite{bakshi_cryptornn-privacy-preserving_2020}.
Both approaches also leak some information about the model and data to the client. The information that needs to be shared is not very valuable though. Both solutions need to share the words that system knows with the client. \cite{podschwadt_classification_2020} additionally needs to share the embedding matrix. This should be fine however, since there are plenty of pretrained embeddings that are publicly available. So, the client does not learn anything new about the model.

When it comes to the security level in bits many solutions opt use 80 bit \cite{al_badawi_high-performance_2018, al_badawi_towards_2020, bourse_fast_2018, nandakumar_towards_2019}. Only  Hesamifard et al. \cite{hesamifard_cryptodl_2017} chooses parameters that provide 128 bit security.

\begin{table}
    \centering
    \caption{Performance impact of the changes to facilitate encrypted execution. * Perplexity per Word (PPW). ** Reported by the authors.  $\dagger$ Values are word error rate (WER)}
    \resizebox{\columnwidth}{!}{
    \begin{tabular}{|c|c|r|r|r|}
        \hline
        Paper                                          & Dataset   & Plain    & Enc.   & Loss \\
        \hline
        Al Badawi \cite{al_badawi_towards_2020}   & MNIST     & -                 & 96\%              & -  \\
        \hline
        Nandakumar \cite{nandakumar_towards_2019} & MNIST     & 96.4\%            & 96\%              & 0.4\% \\
        \hline
         Bourse et al. \cite{bourse_fast_2018} & MNIST   & 96.75\% & 96.43\%            & 0.3 \% \\
        \hline
         Chabanne et al. \cite{chabanne_privacy-preserving_2017} & MNIST   & 99.59\% & 99.30\%            & 0.3 \% \\
        \hline
         Faster Cryptonets \cite{chou_faster_2018} & MNIST   & 99.13\% & 99.17\%            & -0.04 \% \\
        \hline
        CHET \cite{dathathri_chet_2019} & MNIST   & 99.3\% & 99.3\%            & 0 \% \\
        \hline
         Cryptonets \cite{dowlin_cryptonets_2016} & MNIST   & - & 99\%            & - \\
        \hline
         CryptoDL \cite{hesamifard_cryptodl_2017} & MNIST   & 99.56\% & 99.52\%            & 0.04\% \\
        \hline
        E2DM \cite{jiang_secure_2018} & MNIST   & - & 98.1\%            & - \\
        \hline
         SHE \cite{lou_she_2019} & MNIST   & - & 99.77\%            & - \\
        \hline
        Brutzkus et al. \cite{brutzkus_low_2019} & MNSIT & - & 98.95\% & - \\
        \hline
        ngraph-he \cite{boemer_ngraph-he_2019} & MNSIT & - & ~99\% & - \\
        \hline

        Al Badawi \cite{al_badawi_towards_2020}   & CIFAR-10  & -                 & 77.5\%            & -  \\
        \hline
        Faster Cryptonets \cite{chou_faster_2018} & CIFAR-10   & 86.76\% & 75.99\%            & 12.41 \% \\
        \hline
         CHET \cite{dathathri_chet_2019} & CIFAR-10   & 84\% & 81.5\%            & 2.9\% \\
        \hline
         CryptoDL \cite{hesamifard_cryptodl_2017} & CIFAR-10   & 94.2\% & 91.4\%            & 2.9\% \\
        \hline
         SHE \cite{lou_she_2019} & CIFAR-10 & - & 74.1\%            & - \\
        \hline
        Brutzkus et al. \cite{brutzkus_low_2019} & CIFAR-10 & - & 98.95\% & - \\
        \hline 
        ngraph-he \cite{boemer_ngraph-he_2019} & CIFAR-10 & - & 62.10\% & - \\
        \hline 
        
         Mihara et al. \cite{mihara_neural_2020} & Iris & 98.05\% & 98.47\%            & -0.4\% \\
        \hline
         
         SHE \cite{lou_she_2019} & ImageNet & - & 69.4\%            & - \\
        \hline
          Podschwadt \& Takabi \cite{podschwadt_classification_2020} & IMDB & - & 86.47\%            & - \\
        \hline
         PrivFT \cite{al_badawi_privft_2019}       & IMDB      & -                 & 91.5\%            & - \\
        \hline 
         PrivFT \cite{al_badawi_privft_2019}       & YELP      & -                 & 96.1\%            & - \\
        \hline
         PrivFT \cite{al_badawi_privft_2019}       & AGNews    & -                 & 92.5\%            & - \\
        \hline
         PrivFT \cite{al_badawi_privft_2019}       & DBPedia   & -                 & 98.8\%            & - \\
        \hline
        SHE \cite{lou_she_2019} & Penn Treebank & - & 89.8*            & 2.1\%** \\
        \hline
         CryptoRNN \cite{bakshi_cryptornn-privacy-preserving_2020} & \cite{ananda_freire_wall-following_nodate}   & 96.2\% & 96.2\% & 0 \% \\
        \hline
         CryptoRNN \cite{bakshi_cryptornn-privacy-preserving_2020} & \cite{torres_sensor_2013}   & 87.7\% & 87.7\%            & 0 \% \\
        \hline
         CryptoRNN \cite{bakshi_cryptornn-privacy-preserving_2020} & \cite{oliver_roesler_eeg_nodate}   & 72.5\% & 72.5\%            & 0 \% \\
        \hline
         CryptoRNN \cite{bakshi_cryptornn-privacy-preserving_2020} & \cite{candanedo_accurate_2016}   & 99.7\% & 99.7\%            & 0 \% \\
        \hline
        Encrypted Speech \cite{zhang_encrypted_2019} & Switchboard & 12.2\%$\dagger$ & 13.5\%$\dagger$            & 10.7\%$\dagger$ \\
        \hline
        Encrypted Speech \cite{zhang_encrypted_2019} & Cortana & 12.9\%$\dagger$ & 14.8\%$\dagger$            & 14.7\%$\dagger$ \\
        \hline
        Orlandi et al \cite{orlandi_oblivious_2007} & \cite{gorman_analysis_1988} & 84.7\% & - & - \\
        \hline
    \end{tabular}
    }
    \label{tab:performance_impact}
\end{table}

\section{Evaluation} \label{sec:evaluation}

\subsection{Tasks and Datasets}

The task and the data used for evaluation is important for comparison. The data has a large influence on the network architecture and influences the encoding scheme. It is easier to encode a fixed set of discrete values than it is to encode a range of continuous values. Images are often used as they exhibit many beneficial characteristics. Many high performing model architectures have been developed for image processing. Images consist of discreet sets of values and are easily resizable  with little information loss. Smaller images mean less data to encrypt and fewer input features which results in fewer computations within the network and less memory required. The dimensions of other types of data are not as easily manipulated. Text for examples, while consisting of discreet values, cannot be resized. Additionally, models designed for text processing are usually more complex, which results in a higher MD. Below we discuss some of the datasets used. For a complete overview of all datasets used see Table \ref{tab:performance_impact}.

Image classification is a the most common task used for evaluation. Among the used datasets, the MNIST hand written digit dataset is the most widely used \cite{dowlin_cryptonets_2016,chabanne_privacy-preserving_2017, hesamifard_cryptodl_2017,lou_she_2019,chou_faster_2018,dathathri_chet_2019,al_badawi_towards_2020,jiang_secure_2018,bourse_fast_2018,nandakumar_towards_2019}. However, demonstrating that an approach works on MNIST is not enough. It might be good for comparison but as an evaluation dataset it has the significant downside of being too simple and too clean. Strategies that work on MNIST do not necessarily work on other datasets. The CIFAR-10 dataset is a far better choice, since it is closer to a real world dataset than MNIST.
Some solutions, \cite{hesamifard_cryptodl_2017,lou_she_2019,chou_faster_2018,dathathri_chet_2019,al_badawi_towards_2020}, additionally evaluate on the CIFAR-10 data. 
SHE \cite{lou_she_2019} is the only solution which tackles the much harder Imagenet dataset.

Four papers \cite{al_badawi_privft_2019,zhang_encrypted_2019,lou_she_2019,podschwadt_classification_2020} look at language data. Zhang et al. \cite{zhang_encrypted_2019} perform text to speech on the Cortana voice assistant and Switchboard dataset and Lou et al. \cite{lou_she_2019} perform next word prediction on Penn Treebank. Podschwadt \& Takabi \cite{podschwadt_classification_2020} and Al Badawi et al. \cite{al_badawi_privft_2019} evaluate their systems on text classification tasks. 

This makes Zhang et al. \cite{zhang_encrypted_2019} and Lou et al. \cite{lou_she_2019} the only two works that do not only do classification. Classification tasks seem to be easier to handle in the encrypted domain. One explanation is that the output of the neural network is easily interpretable and of comparatively low dimension. As can be seen in Zhang et al. \cite{zhang_encrypted_2019} the interpretation of the network output requires much more computation on the client side. 

\subsection{Resource Requirements}

High resource requirements are shared among all solutions. Comparing their runtime is difficult, since some experiments were performed on a laptop and others on big servers with up to a hundred CPU cores. That is why we only highlight a few noteworthy runtimes. Solutions are more comparable in terms of memory consumption and communication overhead. However, memory requirements are not always reported in the papers. Authors often only report the configuration of their test system. This gives us at least an upper bound on the memory requirement. Unfortunately, not all solutions report their communication overhead either. A complete overview of the memory and communication requirements can be found in Table \ref{tab:ressources}.

\textbf{Communication:}
We can see that batched solutions for MNIST \cite{dowlin_cryptonets_2016, hesamifard_cryptodl_2017, chou_faster_2018} need to transfer about 330-410MB. Due to their packing scheme E2DM \cite{jiang_secure_2018} only needs to transfer 17.4MB. This should be similar for solutions using ciphertext packing. 
As the datastes grow in complexity so does the size of the data. On CIFAR-10 the ciphertexts are much larger except for SHE \cite{lou_she_2019} which only needs 160MB.  Hesamifard et al. \cite{hesamifard_cryptodl_2017} and Faster Crypotnets \cite{chou_faster_2018} need 1,803MB and 1,160MB respectively. To encrypt a single ImageNet image, SHE \cite{lou_she_2019} needs 7.7GB. On medical data of similar size, in  Faster Cryptonets \cite{chou_faster_2018}, the client needs to transfer 789.2GB. Podschwadt \& Takabi \cite{podschwadt_classification_2020} do not use ciphertext packing and the client does need to transfer encrypted vector representations. Initially the client needs to send 14 GB. On top of that there are a number of interactive phases during computation, to remove the noise, during evaluation of the network, which sum up to a total of 15.2 GB.
When training a model on encrypted data the client needs to send the entire training data in encrypted form. On the Youtube Spam Collection, with less than 2000 instance, the encrypted training data in PrivFT \cite{al_badawi_privft_2019} comes in at roughly 550 GB. We can clearly see that the communication overhead for solutions with ciphertext packing is significantly smaller than for solutions without. Solutions with batching can reach pretty low amortized overhead. It is questionable though if that would be beneficial in a real-world setting. It seems much more likely that a client has a few instances for processing instead of the thousands that are required to make the best use out of the naive batching schemes.

\textbf{Memory:}
The size of the ciphertexts is not only an issue in transmission but also has an impact on the memory requirements of the server. Just like with communication the overhead of solutions with ciphertext packing require less memory than those without. Since cloud computing and pay as you use models make machines with hundreds of gigabytes of RAM relatively easily accessible, the memory consumption is only a secondary issue. But if we consider hardware acceleration, such as GPUs, memory becomes an issue again. GPU memory is typically much smaller and can become a bottle neck. Therefore, looking at the memory requirements of these systems can give us a good idea how much they can benefit from hardware acceleration. 
Many solutions \cite{dowlin_cryptonets_2016, hesamifard_cryptodl_2017,brutzkus_low_2019,chou_faster_2018, podschwadt_classification_2020} can be run on commodity hardware, requiring less than 48GB of RAM. As the datasets get more complex the models also become more complex. The Faster Cryptonets \cite{chou_faster_2018} CIFAR-10 model needs 1433.6 GB of memory. Al Badawi et al. \cite{al_badawi_towards_2020} in comparison only needs 187GB. To use GPUs, which have only 16GB of memory, the authors resort to performing subtasks and swapping data in and out of GPU memory.
In comparison training takes less memory. The main difference to the communication overhead is that while all the training data needs to be transmitted, it does not need to be in memory all at once. This is why \cite{al_badawi_privft_2019, nandakumar_towards_2019} can train on 250GB or less. 

\textbf{Runtime:}
As mentioned earlier runtimes are much harder to compare. There are a few we want to highlight. CryptoNets \cite{dowlin_cryptonets_2016} was the first HE based solution on MNIST and takes 250 seconds. Brutzkus et al. \cite{brutzkus_low_2019}, using the same model as CryptoNets \cite{dowlin_cryptonets_2016}, managed to achieve a 2.2s runtime mainly through ciphertext packing. 
As expected, larger datasets and models take longer to run. The fastest CIFAR-10 model is optimized by the CHET compiler \cite{dathathri_chet_2019}. It takes only 164.7s. Which is faster than Al Badawi et al. \cite{al_badawi_towards_2020} model executed on GPUs. On a single GPU the model takes 553 seconds and 304 seconds on four GPUs in parallel. CHET \cite{dathathri_chet_2019} wins out here due to the use of ciphertext packing, which \cite{al_badawi_towards_2020} does not use. But the authors also benefit from the compiler optimizations, which we can see when compared to Brutzkus et al. \cite{brutzkus_low_2019}. Their CIFAR-10 model takes 711 seconds, despite using specialized ciphertext packing strategies. Solutions that do not use ciphertext packing \cite{hesamifard_cryptodl_2017, chou_faster_2018, lou_she_2019} take much longer with 3300-1200 seconds to evaluate the CIFAR-10 model. This is with the downside of that SHE \cite{lou_she_2019} does not support batch processing. It is worth noting that the model of SHE \cite{lou_she_2019} using bootstrapping takes 42.5 million seconds. 

Training on encrypted data is even slower than inference. \cite{nandakumar_towards_2019} need 40 minutes per minibatch on MNIST, with the use of bootstrapping and SIMD batching. \cite{mihara_neural_2020} proposes a different encoding. Using that, it takes only 29 hours to train for 400 epochs. But this is evaluated on a tiny dataset. \cite{al_badawi_privft_2019} uses a more realistic YouTube spam dataset. Here training takes 120 hours using the GPU implementation on 8 GPUs in parallel.
Training on encrypted data does not seem very practical at this point. The resource requirements are almost prohibitively large, especially when coupled with bootstrapping. Bootstrapping is almost a necessity for training due to the noise growth in the encrypted data. If training is performed without bootstrapping either the number of epochs is severely limited or other options for noise removal need to be used, such as interactive phases. 

\subsection{Performance Impact}

All of the approaches compared in this work make changes to the neural network architecture to facilitate execution over encrypted data. As discussed earlier these changes often include different activation functions than those that are used on plain data. But these changes often come at a price in network performance. For example, polynomial activation functions are not ideal from a ML perspective and a concession to the limitations of HE schemes. In Table \ref{tab:performance_impact} we summarize the performance of the models evaluated in the papers. The \textit{Enc.} column shows the performance of models that can be run on plain data, meaning all the operations have been changed to HE friendly alternatives and all changes to reduce complexity have been made. The performance of these models usually does not differ between actual encrypted execution or execution on plaintexts. With the correct crypto parameters, the result is the same.
The \textit{Plain} column gives the performance of the model without any changes to accommodate encrypted execution. This is not supposed to be state of the art, but it serves to show what the model architecture could achieve with standard ML best practices. As can been from the table this information is often missing. The \textit{Loss} column states the relative drop in performance between standard ML model and the model capable of encrypted execution.

Relative loss is an important metric to judge the quality of a PPML solution. Only reporting the result of the HE friendly or encrypted model on the task is not sufficient as it does not capture any performance sacrifices made in order to enable encrypted execution. On the other hand, comparing the performance of the HE friendly model only to the state of the art (if the state of the art performance is achieved using a different architecture) is interesting, but it does not capture the entire picture. State of the art models are often vastly more complex than models that can be run on encrypted data. 
Ideal is a three-way comparison involving state of the art performance on the task, plaintext model and HE friendly model. Reporting performance like this allows for a better comparison of approaches, independent of the task and dataset. 

A large part of the work has been done on image classification and most authors use the MNIST dataset as one, or even, their only evaluation task. However, the relative loss on the MNIST dataset is usually lower than on more complex datasets, such as CIFAR-10. This is not surprising since MNIST is a very simple and clean dataset. Techniques that work on MINST do not necessarily transfer to other datasets. Despite this weakness six of the works considered here only use MNIST. In the field of adversarial examples and defenses, Carlini et al. recommend to not only evaluate on MNIST for this reason. 
We believe a similar recommendation should be made for PPML.

\section{Challenges and Directions} \label{sec:challenges}


As discussed in the previous sections there are still many hurdles to make, HE based PPML practical. In this section we summarize where the challenges lie and discuss potential directions and approaches. Currently there are a handful of libraries available that support one or more encryption schemes. But even if libraries support the same scheme, they are not interoperable. Ciphertexts or keys do not have a standardized format. The same is true for the encryption parameters. The interpretation of what a parameter means can differ between libraries. In the past few years there has been an effort to standardize HE by the community \cite{noauthor_homomorphic_nodate}. This standardization  effort \cite{albrecht_homomorphic_2018} includes an overview of the security and secure parameter recommendations, potential applications and design considerations. It is a good start but at the moment developing solutions locks one into using a particular library. 

Choosing a library can be difficult. In general, the usability of HE is bad. Without in-depth knowledge about HE it is nearly impossible to develop a well performing and secure PPML solution. Developers of PPML need to have expertise in both ML and security. PPML, using HE has not been widely adopted by the ML community at large, likely to the high barrier of entry that is HE and the lack of user friendly tools. On the other hand, security researches often lack ML knowledge. The ML community has done an excellent job of making user friendly tools that allow for fast development and easy entry. 
There should be an effort to create a shared community of ML specialists and crypto experts that drive the development of HE based PPML. Some work has been done on easing the entry from both sides by using compilers such as \cite{boemer_ngraph-he_2019, dathathri_chet_2019}. Compilers not only make it easier to get started, but they can also greatly improve performance by applying optimizations. This can help speeding up execution on encrypted data. But the models are designed with plain text execution in mind. Things that are fast and efficient on plaintext are not always the best solution on encrypted data. A more holistic approach should be taken, that takes the entire ML pipeline into consideration. Much of the work focuses on running existing model architectures on encrypted data. This is often done by making small changes to the model architecture to make it HE compliant and training the model from scratch. Ideally the compiler suite would be able to perform all these transformations automatically while being able to optimize the process end-to-end. Most of the ML techniques used in the field are small adaptions of techniques that work on plaintext. But developing solutions specifically for encrypted execution might require new types of networks, activation functions or optimizers.
Activations are a bottleneck a in neural networks using HE.
These non-linear functions are necessary for the network to perform well. But on encrypted data they greatly impact the noise budget. Multiple alternatives have been proposed but most of them make training the model harder, often due to their unbounded and non-monotonic derivatives. Other solutions that do not impact training such as replacement after training or lookup tables suffer from a loss in model performance. Polynomials are an adequate replacement for traditional activation functions. The question if there are activation functions that perform well, both on encrypted and plain data is still open and worth investigating. 

The need to for HE specific algorithms and adaptions is especially apparent when it comes to training. Training on encrypted data is very limited. One of the big reasons is the noise build up and the resulting need for bootstrapping or interactive noise removal. Solutions that use neither are limited to using training hyper parameters, such as batchsize and number of epochs, that are sub-optimal. Approaches that do use bootstrapping are very slow and computational expensive. There are multiple ways to address the issue. On the one hand faster bootstrapping would remove some of the computational overhead that comes with training on encrypted data and standard training parameters can be used. On the other hand, having a training algorithm that works with the parameters required by HE would eliminate the need for bootstrapping or interactive noise removal and make training much more feasible.

Performance is not only an issue during training. Scaling up to larger, real-world dataset is costly. Models on the CIFAR-10 dataset have can be evaluated within reasonable time. But larger models and larger inputs still take hours to compute. Improvements in latency and memory consumption have been due to ciphertext packing. Without it scaling up to real-world data is impractical. Packing can reduce the memory requirement by an order of magnitude. Models on the CIFAR-10 dataset, that do not use packing, require 100s of GB of memory, making scaling up to datasets like ImageNet near impossible. Another potential source of increased performance are hardware accelerators like GPUs or FPGAs. \cite{al_badawi_towards_2020} have successfully run models using HE on GPUs, but memory constraints are more severe. Even high-end GPUs do not have more than 50GB of memory. The memory limitation of GPUs calls for new encoding and packing schemes for efficient execution. Research on GPU aided HE is not limited to ML applications. For example, Geraldo et al. \cite{alves_faster_nodate} propose an implementation of BFV accelerated by GPUs. Their results show an up to 5 times speed up in homomorphic operations. Increasing the efficiency of neural networks can also help reducing the memory and runtime requirements of PPML. A lot of work has been done on running ML applications on resource constraint devices like phones. Lessons learned in that area can be transferable to PPML. \cite{chou_faster_2018} have looked at reducing the number of computations in the network through pruning. Helbitz \& Avidan \cite{helbitz_reducing_2021} tackle the computational overhead of privacy preserving neural networks by reducing the ReLU count in the network. They show that activations can be grouped together and the output of one activation can be used for all in the group. The accuracy impact on this grouping is low while it can provide about 30\% speedup. They test their system using 2 an 3 PC and not approaches that rely purely on HE. If a similar speedup and can be observed on HE only systems or if it is even applicable to other activations than ReLU still needs to be investigated.

Comparing the speedup and performance of different solutions is often not straight forward. As mentioned in previous sections comparing runtimes is an issue due to different hardware being used. A metric that is hardware independent is the memory requirement. But we find that only some solutions report their memory requirements. We propose a set of evaluation criteria that make comparison of approaches easier and more meaningful. 
1.) the memory requirements 2.) state of the art performance for the task 3.) performance of the model on encrypted data 4.) performance of a model that uses the same number of layers and neuron as the model on encrypted data but also uses standard activation functions. This would give an indication of how much performance loss, if any, privacy preservation adds.


\bibliographystyle{IEEEtran}
\bibliography{references}

\end{document}